\documentclass[aps,prl,twocolumn,]{revtex4}
\usepackage{amsmath}
\usepackage{amsfonts}
\usepackage{amssymb}
\usepackage{bm}
\usepackage{color}
\usepackage{graphicx}

\def\br{\bm r}
\def\bR{\bm r}

\begin{document}
\title{Extreme events in the dispersions of two neighboring particles under the influence of fluid turbulence}
\author{R. Scatamacchia$^{1,2}$, L. Biferale$^{1}$, F. Toschi$^{2,3}$}
\affiliation{$^1$Dept. Physics \& INFN, Univ. Tor Vergata, Via della
  Ricerca Scientifica 1, 00133 Rome, Italy.\\
$^2$Dept. Physics, Eindhoven Univ. of Technology, 5600 MB
  Eindhoven, The Netherlands.\\
$^3$Dept. Mathematics and Computer Science Eindhoven
  Univ. of Technology, 5600 MB Eindhoven The Netherlands \& CNR-IAC,
  Via dei Taurini 19, 00185 Rome, Italy.}  
\date{\today}

\begin{abstract} We present a numerical study of
  two-particle dispersion from point-sources in $3D$ incompressible
  Homogeneous and Isotropic turbulence, at Reynolds number $Re \simeq
  300$. Tracer particles are emitted in bunches 
from localized sources smaller than the Kolmogorov scale. 
 We report the first quantitative evidence, supported by an  unprecedented statistics, of the deviations  of
  relative dispersion from Richardson's picture. Deviations are
  due to extreme events of pairs separating much faster than average,
  and of pairs remaining close for long times. The two classes of
  events are the fingerprint of complete different physics, the former
  being dominated by inertial subrange and large-scale fluctuations,
  while the latter by the dissipation subrange.
  A comparison of relative separation in surrogate white-in-time
  velocity field, with correct viscous-, inertial- and integral-scale
  properties allows us to assess the importance of temporal
  correlations along tracer trajectories.
\end{abstract}

\maketitle
\begin{figure}
 \includegraphics[scale=0.6,draft=false]{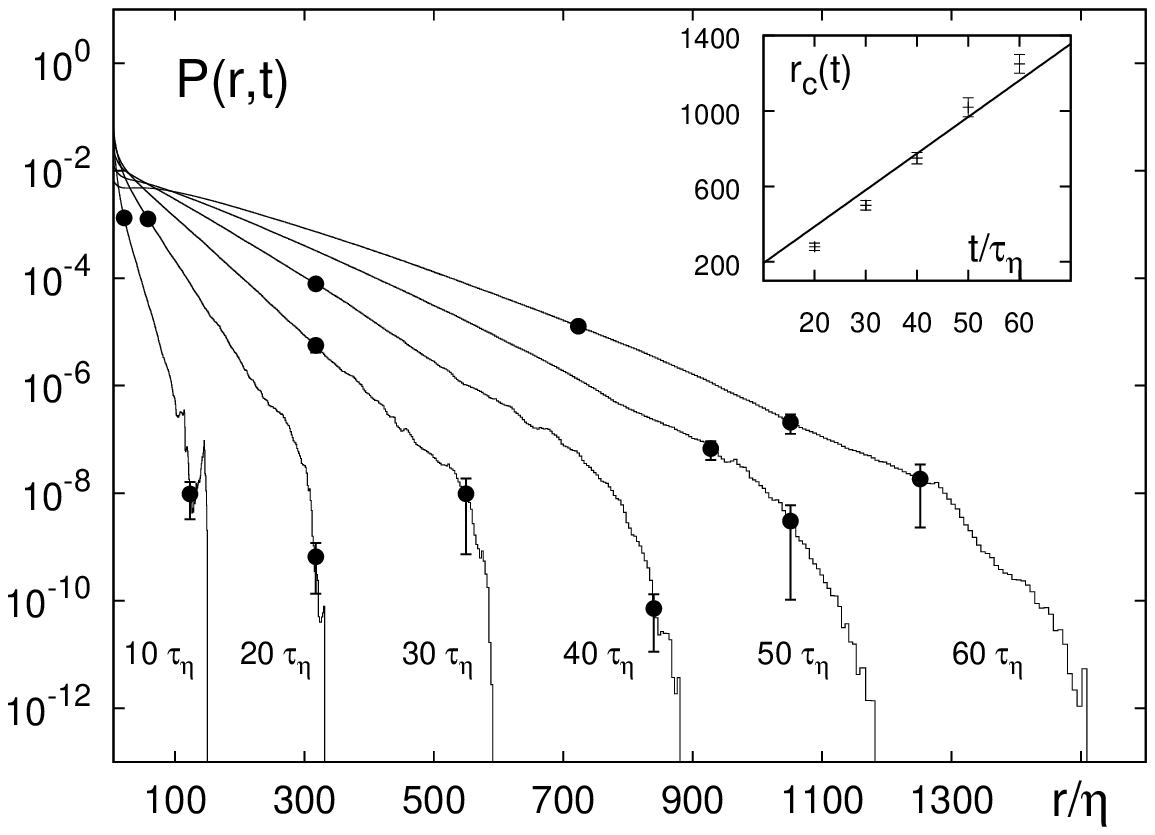}
 \includegraphics[scale=0.6,draft=false]{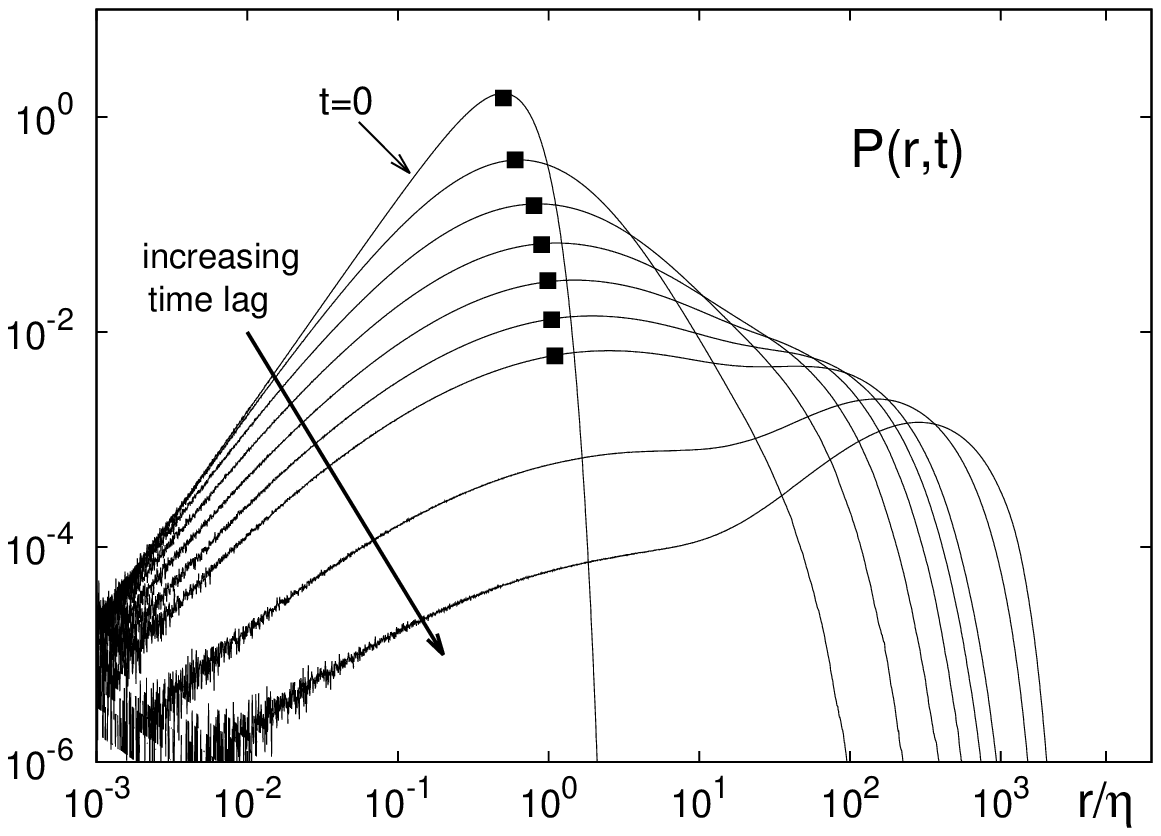}
 \caption{Top: Log-lin plot of $P(r,t)$ at different times after the
   emission. For selected values of $r$, we show error bars, estimated
   from the statistical spread of different runs. Inset: evolution of
   the cut-off scale $r_c(t)$, the continuous line represents the ballistic motion $\propto v_{rms} t$. 
Bottom: log-log plot of $P(r,t)$ for
   $t=(10,20,30,40,50,60,70,90,120)\tau_\eta$.  Black squares indicates the peak observed for small separations.}
  \label{fig:2}
\end{figure}

\begin{figure}
\includegraphics[scale=0.6,draft=false]{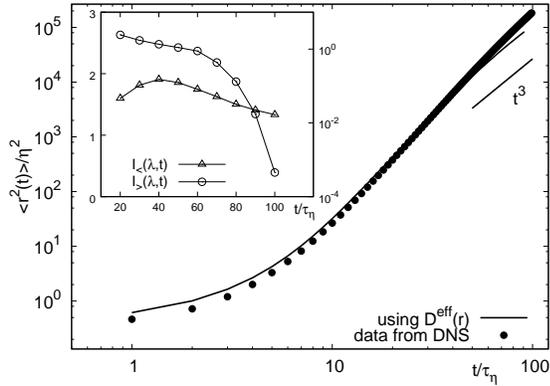}
\caption{Log-log plot of $\langle
  r^2(t)\rangle$ from DNS data, and from the diffusive
  evolution with eddy diffusivity (\ref{eq:eff}). Inset: time
  evolution of the relative probability to observe a large excursion,
  $I_>(\lambda, t)$ (right $y$-scale), or small excursion
  $I_<(\lambda,t)$ (left $y$-scale) for $\lambda=3$.}
\label{fig:4}
\end{figure}

\begin{figure}
  \includegraphics[scale=0.6,draft=false]{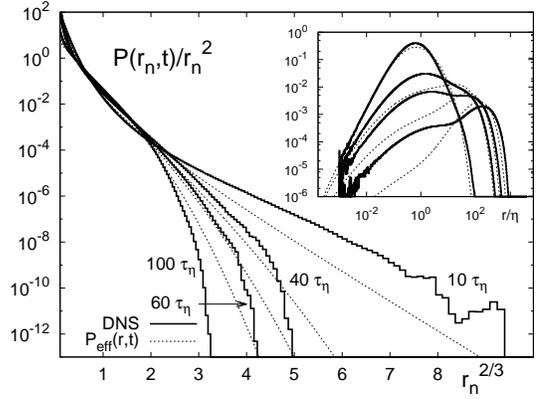}
  \caption{Lin-log plot of $P_{eff}(r,t)$ as obtained from the
    integration of (\ref{eq:pdfRich}) with $D^{eff}(r)$ (dashed) and
    the DNS data (solid line). Inset:log-log plot to highlight the
    slowest events (left tail).}
  \label{fig:5}
\end{figure}

\begin{figure}
  \includegraphics[scale=0.6,draft=false]{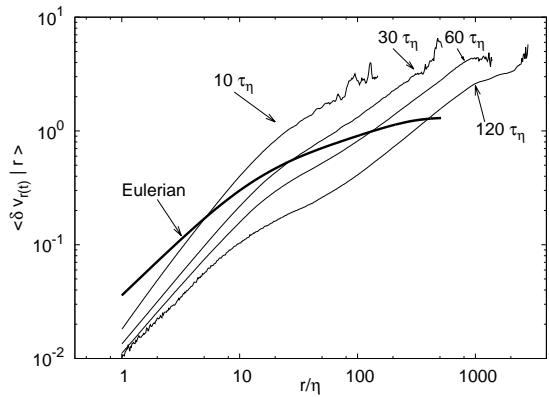}
  \caption{Comparison between conditional Lagrangian (thin lines) and
    Eulerian (thick line) velocity increment moment.}
  \label{fig:6}
\end{figure}

 
The relative separation of pairs of fluid particles by turbulence has
been first addressed by L.F. Richardson \cite{rich} in a pioneering
paper in 1926 (see \cite{SC09,B12} for recent reviews). The main
question is simple and fundamental: given a pair of particles
released, at time $t_0$, at a small separation $\br_0$ 
(smaller of the Kolmogorov dissipative scale, $\eta$) what is the
probability to find the pair at any given distance, $r$, at a later time
$t$? 
%
In the case of isotropic and homogeneous turbulence (HIT) the
probability density function (PDF), $P(\bR,t |\bR_0,t_0)$, of pair
separations depends on the amplitude of $\bR(t)$ only.  Moreover,
asymptotically it should become independent of the initial condition,
$P(r,t)$.\\ The knowledge of this distribution is of utmost importance for many
applied studies in geophysics \cite{Su71,Olli05,Lacasce10}, being connected also to risk-evaluations in environmental disasters.
 Moreover, it  constitutes a highly non trivial
statistical problem, due to its intrinsic non-stationary nature which
connects Lagrangian properties, i.e. velocity fluctuations evaluated
at the particle positions, along their whole past history
\cite{FGVrev}. Richardson proposed to model the particle separation,
at inertial subrange distances $\eta \ll r \ll L_0$, as a diffusive
process characterized by an effective turbulent diffusivity,
$D_{Ric}(r)=\frac{1}{2}\frac{d \langle r^2 \rangle}{dt}$, estimated
empirically to follow a $4/3$-law:
\begin{equation}
D_{Ric}(r) \sim \beta r^{4/3}\,.
\label{eq:eddy0}
\end{equation}
Here $L_0$ is the large-scale of the flow and $\beta=k_0
\epsilon^{1/3}$, where $k_0$ is a dimensionless constant and
$\epsilon$ is the turbulent kinetic energy dissipation. It is easy 
Richardson's work with the  Kolmogorov 1941 theory
\cite{Fr95} by
following the dimensional estimate \cite{FGVrev,So99}:
\begin{equation}
D_{Ric}(r) \sim \tau(r) \langle (\delta_r v)^2 \rangle\,,
\label{eq:eddyEul}
\end{equation}
where $\tau(r) \sim \epsilon^{-1/3}r^{2/3}$ is the eddy-turn-over-time
at scale $r$ and $\langle (\delta_r v)^2 \rangle =C_0 \epsilon^{2/3}
r^{2/3}$ is the second order Eulerian longitudinal structure
function. The resulting long time growth of the mean
squared separation is:
\begin{equation}
\langle r^2(t) \rangle = g \epsilon t^3\,,
\label{eq:r2} 
\end{equation}
$g$ being the Richardson constant uniquely determined in terms of
$k_0$ \cite{OM00,BS02a,BBCDLT05,S08}.
Since \cite{rich}, many experimental and numerical studies
\cite{JPT99,NV03,BBCDLT05,OXBB06} have focused on the subject,
including also extensions to the case of particles with inertia
\cite{BBLST10,D08,FH08}. There are not doubts that Richardson's picture captures some important 
features of turbulent dispersion, in particular concerning events with a typical separation of the order 
of the mean. On the other hand, fundamental questions exist about the capability of the 
 theory to correctly predict also extremal events, 
 i.e. those pairs with separation much larger/smaller than  $\langle r^2(t) \rangle^{1/2}$. Such events are very 
difficult to measure, due to their low probability in absolute terms.  Richardson's
approach can be reinterpreted as the evolution of tracers in a
stochastic Gaussian, homogeneous and isotropic velocity field,
delta-correlated in time, with two-point correlation
\cite{K68}:
$\langle \delta_{r}v_i(t)\delta_{r}v_j(t')\rangle =\delta(t-t')[ D_{\parallel}(r) \hat r_i \hat r_j + D_{\perp}(r) (\delta_{ij} - \hat r_i \hat r_j)]\,,\nonumber$
where the longitudinal and transverse second order structure
functions $D_{\parallel}(r)$ and $D_{\perp}(r)$ are such that
$D_{\perp}(r) = D_{\parallel}(r) + 1/2\, r \partial_r
D_{\parallel}(r)$, by incompressibility. Under this assumption, the
evolution of 
$P(r,t)$ is closed and local: \cite{K68,FGVrev}:
\begin{equation}
\label{eq:Rich}
\partial_t P(r,t) = r^{-2} \partial_r r^2 D_{\parallel}(r) \partial_r P(r,t)\,.
\end{equation}
\begin{figure}
\includegraphics[scale=0.5,draft=false]{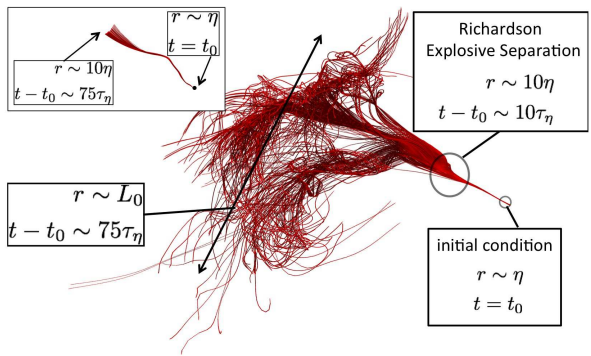}
\vspace{-2cm}
  \caption{Typical time history up to $ t = 75 \tau_\eta$ of a bunch
    emitted from a source of size $\sim \eta$.
  Inset: time
    history for the same duration of a bunch emitted in a different
    location and which does not separate. DNS are performed on a cubic
    fully periodic grid at $1024^3$ collocation points with
    a pseudospectral code, at a Reynolds-Taylor number, $Re_\lambda \sim 300$. For further details on the numerics see
    \cite{BBCDLT05}.
}
  \label{fig:1}
\end{figure}
\noindent Whenever the eddy-diffusivity  has a power-law
behavior, $D_{\parallel}(r) =D_0 r^{\xi}$ with $ 0 \le \xi < 2$, the
above equation with initial condition $P(r,t_0)\propto \delta(r-r_0)$
can be solved in terms of special functions \cite{BL98,Be06}. So
doing, it admits the celebrated asymptotic large time form
(independent of $r_0$):
\begin{equation}
  \label{eq:pdfRich}
  P_{Ric}(r,t) \propto  \frac{r^2}{\langle r^2(t)\rangle^{\frac{3}{2}}} \exp\left\{-b\left(\frac{r}{\langle r^2(t)\rangle^{\frac{1}{2}}}\right)^{2-\xi}\right\}\,.
\end{equation}
Here $b$ is a constant uniquely determined in terms of $D_0$
\cite{BL98}, which plays the same role of $\beta$ in
eq.~(\ref{eq:eddy0}). In such {\it idealized} scaling scenario, tracer
pairs separate in a {\it explosive} way forgetting their initial
separation\,: $
\langle r^2(t) \rangle \propto  t^{2/(2-\xi)},  
$
which reproduces Richardson's expression for $\xi=4/3$.\\ When the
turbulent flow is differentiable, i.e. $\xi =2$, the PDF takes the
log-normal form, $P(r,t|r_0,0) \propto
\exp{\left\{-(\log(r/r_0)-\lambda t)^2/(2 \Delta\ t)\right\}}$,
where $\lambda$ is the first Lyapunov exponent and $\Delta$ is
connected to fluctuations of the strain matrix \cite{FGVrev}. In
the latter case, particles separate exponentially and the memory of
the initial separation, $r_0$, remains at all times. The rate of
separation is strongly fluctuating from point-to-point and from
time-to-time being connected to the fluctuations of the Lyapunov
exponents \cite{BBBCMT06,FGVrev}.
\\The behavior of particle
pairs in real flows can deviate from Richardson
distribution because of different reasons. The four most important
are: (i) temporal correlations of the underlying velocity fluid
\cite{K68,So99,IPZ10}; (ii) non-Gaussian velocity fluctuations; (iii)
ultraviolet (UV) effects induced by the dissipation subrange, and
(iv) infrared (IR) effects induced by the large-scale cut-off. These
last two are of course connected to finite Reynolds effects
\cite{SYH08}.\\
%
The goal of
this Letter is twofold. First, we want to understand and quantify the
rate of occurrence of {\it rare} extreme events, i.e., of pairs that
separate much more or much less than $\langle r^2(t)\rangle^{1/2}$. 
Second, we aim to
assess the importance of temporal correlations for the pair statistics
both in general and for extreme events, in particular.\\ We have
performed a series of Direct Numerical Simulations (DNS) of HIT
seeding the fluid with bunches of tracers, emitted in different
locations, to reduce spatial correlations. Each bunch is emitted
within a small region of size $\sim \eta$, in puffs of $2 \times
10^3$ tracer particles each. In a single run, there are $256$ of such
sources, emitting about $200$ puffs with a frequency comparable
with the inverse of the Kolmogorov time. We performed
$10$ different runs, following a total of $4 \times 10^{11}$ pairs,
reaching an unprecedented statistics.  In Fig. (\ref{fig:1}) we
illustrate the complexity of the problem. We first notice the abrupt
transition in the particle dispersion occurring at a time about $\sim 10 \,\tau_\eta$ after the emission
when most of the pairs reaches a
relative distance of the order of $\sim 10
\eta$, and the beginning of an  explosive separation {\it \'a la } Richardson is
observed. At at later time, there are many pairs with relative separations of
the order of the box size $ \sim 1000 \eta$, even
though the mean separation is much smaller at those time lags. On the contrary, in the
inset we show an example of a bunch with an anomalous history, 
due to tracers that travel close -at mutual distance of the order of
$\eta$ -, for very long times. This happens when pairs are injected in
a space location where the underlying fluid has a small local
stretching rate. Nevertheless, once the pairs reach inertial subrange
separations, the bunch rapidly expands forgetting its initial delay
and recovers at large times a spread distribution (not shown
here).\\ To quantify this phenomenology, we show in Fig.~(\ref{fig:2})
the right and left tails of  $P(r,t)$ at different time lags,
averaged over all emissions and over all point sources. The top panel
shows the fastest separation events, with a clear exponential-like
tail plus a sharp drop at a cut-off separation, $r_c(t)$, that evolves
in time. The {\it cut-off scale} $r_c(t)$ is connected to pairs which
are able to separate ``very fast'', and it is a clear indication of
the change in the physics governing large excursions. It indicates the 
presence of a finite
maximal propagation velocity: it is the signature of tracer pairs
experiencing a persistently high relative velocity, which must be 
limited by the root-mean-squared velocity, $v_{rms}$ \cite{OT06}. To
support this statement  we show in the inset the evolution of
$r_c(t)$ which is in good agreement with a linear behavior obtained
using $v_{rms}$ as traveling speed. Events beyond the cut-off scale  are
{\it rare}, and can be detected with high-statistics
only.\\
The opposite limit of ``very slow'' events is also 
remarkable (see bottom panel of Fig.~\ref{fig:2}). Here we
observe a bi-modal shape for $P(r,t)$ at almost all times:
the left tail of pairs with mutual distance $r<\eta$ remains populated
 for time up to $\sim 60-70 \tau_\eta$, which is of the order of
the large-scale eddy-turn-over-time, $T_L$. Pairs
emitted in regions with a {\it small} stretching rate tend to stay
together, reaching only very late the inertial subrange separations,
and thus never experiencing a Richardson-like dispersion. These pairs
are governed by a log-normal distribution. This is a non-trivial
result in tracers dispersion, that can not be brought back to
small-scale clustering effects as those observed in the dynamics of
inertial particles \cite{BBBCCLMT06,BST10}.\\
\begin{figure}
  \includegraphics[scale=0.6,draft=false]{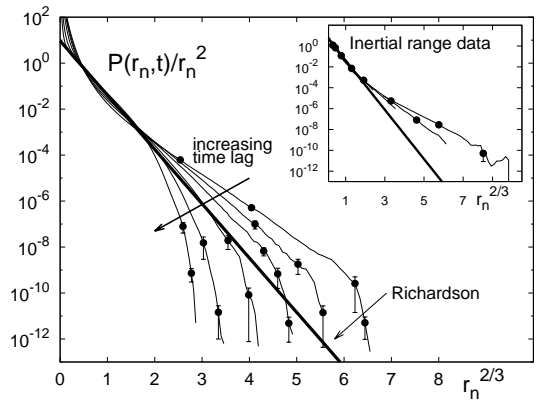}
  \caption{Log-lin plot of $P(r_n,t)$ versus the rescaled variable
    $r_n$ (see text) for $t=(20,30,40,60,90,120)\tau_\eta$. The
    distribution $P(r_n,t)$ has been divided by a factor $r_n^2$ to
    highlight the large separation range. Richardson prediction
    (\ref{eq:pdfRich}) becomes time-independent if rescaled in this
    way (solid curve). Inset: same PDFs plotted only for separations
    $r_n$ that, at any time lag, belong to the inertial subrange.}
  \label{fig:3}
\end{figure}
In
Fig.~(\ref{fig:3}) we plot the same data of Fig. (\ref{fig:2}) but rescaled in
terms of the variable $r_n(t) = r(t)/\langle r^2(t)\rangle^{1/2}$, and
compared against the asymptotic prediction (\ref{eq:pdfRich}). Here,
the deviations from Richardon's prediction becomes clear, showing
evident discrepancies at large scales for all times. A more stringent
test is obtained by showing these same PDFs  but restricted to
the scales in the inertial subrange, $30 \eta < r < 300 \eta$
(inset). Clearly (\ref{eq:pdfRich}) is not 
satisfied. Previous studies could access events up to $r/\langle
r^2(t)\rangle^{1/2} <3$ only (see
\cite{SC09}).  Our study improves of four-five order of magnitudes (in probability) 
the intensities of detectable events, thus allowing to highlight strong deviations from Richardon's shape. 
 Large discrepancies can be measured also on the left
tails of $P(r,t)$, associated to very slow separating pairs (see also below).\\ Such departures from the {\it ideal} self-similar Richardson
distribution needs to be better quantified, either in terms of finite
Reynolds effects (break-up of self-similarity of the turbulent eddy
diffusivity) or in terms of the neglected temporal correlations, or
both.\\ To assess the importance of the former, we
have integrated the Richardson equation
(\ref{eq:Rich}) using an effective eddy-diffusivity $D^{eff}(r)$ which
improves (\ref{eq:eddyEul}) by including UV and IR cut-offs. Depending
whether the separation distance falls in the viscous, inertial or
integral range of scales, we must then have\,:
\begin{equation}
\label{eq:eff}
\begin{cases}
  D^{eff}_{\parallel}(r) \sim r^2  \qquad\quad r \ll \eta  \\
  D^{eff}_{\parallel}(r) \sim r^{4/3}  \qquad \eta \ll r \ll L_0 \\
  D^{eff}_{\parallel}(r) \sim const. \,\,\,\,\,\, r \gg L_0.
\end{cases}
\end{equation}
A widely used fitting formula that reproduces well the Eulerian data,
and that matches the expected UV and IR scaling for both $\tau(r)$ and
$\langle (\delta_r v)^2\rangle$, is obtained by following \cite{Me96}\,:
\begin{equation}
\label{eq:fitmeneveau}
  \langle (\delta_r v)^2\rangle =  c_0\frac{r^2}{((r/\eta)^2 + c_1)^{(2/3)}}  \left[1+c_2 (\frac{r}{L_0})^2\right]^{-1/3}\,
\end{equation}
supplemented with a similar expression for the eddy turn over time: $ 
 \tau(r) = \frac{\tau_\eta}{((r/\eta)^2+c_1)^{-1/3}}  \left[1+d_2(r/L_0)^2\right]^{-2/3}
$. The dimensionless parameters $c_0, c_1, c_2$ are extracted from
the Eulerian statistics, while the parameter $d_2$ is fixed such as to
correctly reproduce the evolution of the mean square separation,
$\langle r^2(t)\rangle$, over a time range $\tau_\eta \le t \le T_L$,
see Fig.~\ref{fig:4}. Note that the hypothesis of Gaussian statistics
still (implicitly) holds in this approach, since the velocity field
distribution is fixed in terms of the second order moment only.\\
  The solution, $P_{eff}(r,t)$, to
the diffusive equation (\ref{eq:Rich}) using $D^{eff}(r)$ is shown in
Fig.~\ref{fig:5} and compared with the DNS data. Despite the excellent agreement 
between $\langle r^2(t) \rangle$ obtained from DNS and the
one obtained using the stochastic model (\ref{eq:Rich}) and
(\ref{eq:eff}), the far tails are still different.
Self-similarity is broken by the introduction of UV and
IR cutoffs in Eqn.~(\ref{eq:eff}), and therefore $P_{eff}(r,t)$ no longer rescales at
different times: this is at variance with Richardson idealized
picture, and clearly goes in the direction of real turbulent flows. 
For large separations the agreement with the DNS
data is qualitatively better but still quantitatively off,
particularly when focusing on the sharp change at the cut-off scale
$r_c(t)$ which is still absent in the evolution given by
(\ref{eq:eff}).  This is a key-point showing that to reproduce the
observed drop in the PDF for large excursions, it is not enough to
impose a saturation of eddy-diffusivity for large $r$.  The behavior
of pair dispersion must then be crucially dependent on the nature of
temporal correlations of the Lagrangian turbulent velocity in the
inertial subrange. Similarly, if we focus on the small separation
tail, $P_{eff}(r,t)$ presents a
slowly evolving peak for $r \le \eta$ but quantitative agreement is
again not satisfactory yet (inset of Fig.~\ref{fig:5}). We
understand such departure as the effects of assuming a Gaussian
velocity statistics underlying the  evolution of $P_{eff}(r,t)$, which is blatantly wrong because of turbulent
small-scale intermittency. To further quantify the
departure of the modified Richardson description from the real data we measured the cumulative
probability to have a couple at large separation $r^*(\lambda)= \lambda
\langle r^2(t)\rangle^{1/2}$ normalized with the same quantity
evaluated from Eqn.~(\ref{eq:pdfRich}) using (\ref{eq:eff}), namely
$I_>(\lambda,t) = \int_{r^*(\lambda)}^{\infty} dr P(r,t)/\int_{r^*}^{\infty} dr
P_{eff}(r,t)$; similarly, to  evaluate the differences for small separation events, we use $r^*(\lambda)= 1/\lambda \langle
r^2(t)\rangle^{1/2}$ and define: $I_<(\lambda,t) =
\int_0^{r^*(\lambda)} dr P(r,t)/\int_0^{r^*} dr
P_{eff}(r,t)$. The results are shown in the inset of
Fig.~\ref{fig:4} for $\lambda=3$. Concerning $I_<(\lambda,t)$, 
it shows that the evolution using $P_{eff}(r,t)$ underestimate of a factor $2$ at small time-lags, $\sim
20 \tau_\eta$ the importance of the small-scale trapping properties, i.e. does not capture the strong intermittency of 
the regions where we have a small stretching rate.  Only for very large times
$\sim 100 \tau_\eta$ the left tail becomes comparable with the real ones. Concerning $I_>(\lambda,t)$ we measure first an
underestimate of large separation events and later a strong overestimate, i.e. the delta-correlated approach behind the evolution of $P_{eff}(r,t)$ does not capture the presence of the cut-off scale $r_c(t)$.\\
\noindent To directly 
quantify the importance of temporal correlations along tracer
trajectories,  we compare in Fig. (\ref{fig:6}) Lagrangian longitudinal
velocity increments conditioned on particle separation distance $r$\,:
\begin{equation}
S_{Lag}(r,t) = \langle (\delta_{r(t)} v_i \cdot \hat r_i) |r(t) = r \rangle\,,
\label{eq:S2lag}
\end{equation}
with its  Eulerian equivalent\,:
\begin{equation}
S_{Eul}(r) = \langle |\delta_{r} v_i\cdot \hat r_i| \rangle\,.
\label{eq:S2eul}
\end{equation}
The figure shows that pairs that have separated fast have also a
typical velocity difference higher than the Eulerian one, measured on
the whole volume without conditioning on the history of
particles. This is in agreement with what found before: intense
separation events feel highly persistent velocity increments along
their time history, and they cannot be trivially estimated starting
from Eulerian statistics. Figure (\ref{fig:6}) also shows the opposite
trend: pairs that have not separated much, at any given time, have
Lagrangian velocity increments smaller than the Eulerian ones. 

In conclusion, we have presented the first numerical study
at very high statistics meant to explore rare events in turbulent dispersion.  Both ``fast'' and ``slow''
separations show an important departure from the
Richardson-like inertial and idealized behavior of
Eqn.~(\ref{eq:pdfRich}).
A step forward toward a better modelization is
obtained by  maintaining the assumption of time
delta-correlation and Gaussian statistics but introducing an improved
effective eddy-diffusivity kernel, which takes correctly into account
both the viscous and integral scale physics. Doing so, a better
qualitative agreement with real data is observed.  To further
progress we will need to relax the Gaussianity assumption for the
small-separation and the delta-correlated assumption for separations
of the order of the integral scale. An attempt following the latter
direction by using a Langevin-process for the evolution of the
relative particles velocity, with different correlation times at
different scales, has been recently proposed by
\cite{Be12}. Alternatively, a memory kernel can be introduced in
order to overcome the presence of unphysical infinite speed events
\cite{IPZ10}, characterizing pairs that move in a delta-correlated
velocity field. For example,
 1d model  with telegraph equations shows
similar {\it maximum} speed events \cite {OT06}.\\
\noindent We thank A.S. Lanotte for collaboration in a early stage of
this work and M. Cencini, J. Bec and R. Benzi for discussions. We
acknowledge the EU COST Action MP0806 ``Particles in turbulence'' and
CINECA (Italy) within the HPC program ISCRA-Class A Project. This
research was supported in part by the Project of Knowledge Innovation
Program (PKIP) of Chinese Academy of Science, Grant No. KJCX2.YW.W10.

\end{document}